\documentclass[aps,prl,twocolumn,superscriptaddress,]{revtex4-1}
\usepackage{amsmath}
\usepackage{amsfonts}
\usepackage{graphicx}
\usepackage{color}
\usepackage{dsfont}
\usepackage{verbatim}
\usepackage{mathtools}
\usepackage[normalem]{ulem}
\usepackage{comment}
\usepackage{stackrel}
\usepackage{bm}
\usepackage[pdftex]{hyperref}
\hypersetup{colorlinks = true, urlcolor = black, linkcolor = black, citecolor = black}

\definecolor{myblue}{rgb}{.93, .93, 1}

\newcommand{\bsub}{\begin{subequations}}
	\newcommand{\esub}{\end{subequations}}

\newcommand{\vex}[1]{\bm{\mathrm{#1}}}

\newcommand{\tr}{\mathsf{Tr}}

\begin{document}
	
\title{Correlation-induced triplet pairing superconductivity in graphene-based moir\'e systems}
	
\author{Yang-Zhi~Chou}\email{yzchou@umd.edu}
\affiliation{Condensed Matter Theory Center and Joint Quantum Institute, Department of Physics, University of Maryland, College Park, Maryland 20742, USA}
	
\author{Fengcheng~Wu}
\affiliation{School of Physics and Technology, Wuhan University, Wuhan 430072, China}
	
\author{Jay D. Sau}
\affiliation{Condensed Matter Theory Center and Joint Quantum Institute, Department of Physics, University of Maryland, College Park, Maryland 20742, USA}

\author{Sankar Das~Sarma}
\affiliation{Condensed Matter Theory Center and Joint Quantum Institute, Department of Physics, University of Maryland, College Park, Maryland 20742, USA}	
\date{\today}
	
\begin{abstract}
Motivated by the possible non-spin-singlet superconductivity in the magic-angle twisted trilayer graphene experiment, we investigate the triplet-pairing superconductivity arising from a correlation-induced spin-fermion model of Dirac fermions with spin, valley and sublattice degrees of freedom. We find that the $f$-wave pairing is favored due to the valley-sublattice structure, and the superconducting state is time-reversal symmetric, fully gapped, and non-topological. With a small in-plane magnetic field, the superconducting state becomes partially polarized, and the transition temperature can be slightly enhanced. Our results apply qualitatively to Dirac fermions for the triplet-pairing superconductivity in graphene-based moir\'e systems, which is fundamentally distinct from triplet superconductivity in $^3$He and ferromagnetic superconductors. 
\end{abstract}
	
\maketitle

\textit{Introduction.--} Since the incipient discovery of correlated insulators and superconductivity in magic-angle twisted bilayer graphene (MATBG) \cite{tbg1,tbg2}, the moir\'e graphene systems continue to uncover exotic phases and excite new ideas \cite{Yankowitz2019,Polshyn2019,Cao2020PRL,Sharpe2019,Lu2019,Kerelsky2019,Jiang2019,Xie2019_spectroscopic,Choi2019,Serlin2020,Park2021flavour,Chen2019signatures,Burg2019,Shen2020correlated,Cao2020tunable,Liu2020tunable,Park2021tunable,Hao2021electric,Cao2021}. 
In particular, magic-angle twisted trilayer graphene (MATTG) \cite{Park2021tunable,Hao2021electric,Cao2021} establishes a second example of the robust superconductivity in the moir\'e graphene systems that reveals a clear Fraunhofer-like oscillation
in a Josephson interference pattern. In addition, the out-of-plane displacement field can modify the band structure significantly, providing a controllable way to tune the Van Hove singularity as well as the superconductivity \cite{Park2021tunable,Hao2021electric,Cao2021}.

A recent experiment in MATTG \cite{Cao2021} demonstrated that the superconducting state survives with a large in-plane magnetic field (until $\sim$10T) that exceeds the Pauli limit for spin-singlet superconductivity, \textit{prima facie} implying a non-spin-singlet superconducting state in MATTG. Remarkably, the experiment also found a non-monotonic superconducting behavior as a function of the applied in-plane magnetic field, which suggests a separate (re-entrant) superconducting phase for magnetic field beyond 8T.

\begin{figure}[t!]
	\includegraphics[width=0.45\textwidth]{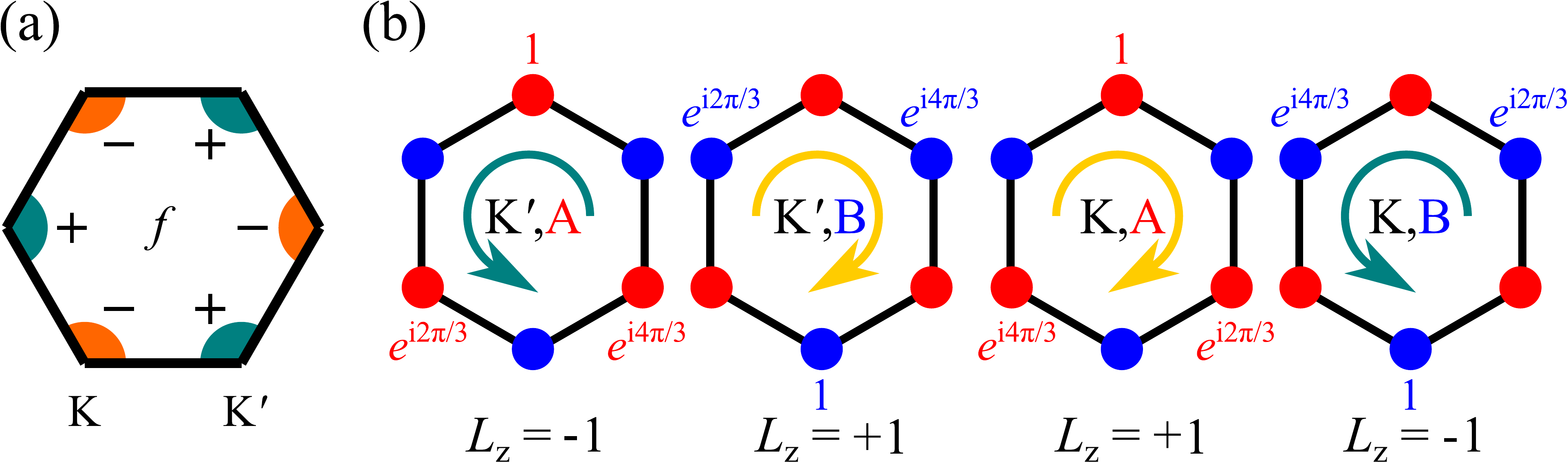}
	\caption{(a) $f$-wave pairing symmetry and the Brillouin zone. The signs indicate the relative phase when a $f$-wave pairing is rotated by $\pi/3$. K and K$'$ denote the valleys.
	(b) Angular momentum and valley-sublattice structure. The electrons carry finite angular momenta depending on the valley and sublattice as depicted in the figure. Therefore, the Cooper pair wavefunction might carry nontrivial $z$-directional angular momentum ($L_z$). A and B denote the sublattices.
	}
	\label{Fig:Sub_pairing}
\end{figure}

In this Letter, we study a spin-fermion model \cite{Monthoux1999,Blagoev1999,Roussev2001,Abanov2001,Chubukov2003,Abanov2003quantum} for Dirac fermions with spin, valley and sublattice degrees of freedom, as a proxy for investigating the possible spin-triplet superconductivity in graphene-based moir\'e systems. 
The idea is that the system is proximate to correlation-induced ferromagnetism even when the long-ranged order is not observed. The fluctuation of such a phase (i.e., spin fluctuation) is captured by the spin-fermion model, and the spin fluctuation generates superconductivity regardless of the details in the band structure.
We establish that the spin-triplet superconducting state is $f$-wave [see Fig.~\ref{Fig:Sub_pairing}(a)] and fully gapped due to the valley-sublattice structure in the Cooper pairs. 
We also discuss the effect of a small in-plane magnetic field and experimental characterization.

We note that the sublattice and valley structures in the pairing states crucially determine the dominating pairing instability, while the moir\'e band structures in twisted graphene systems inherit Dirac fermions with internal degrees of freedom (e.g., spin, valley, sublattice) from monolayer graphene. Therefore, we anticipate that the $f$-wave is generically the dominating pairing symmetry for spin-triplet superconductivity in the graphene based systems.

\textit{Model.--} The spin-fermion model \cite{Monthoux1999,Blagoev1999,Roussev2001,Abanov2001,Chubukov2003,Abanov2003quantum} describes that the low-energy itinerant electrons interact with the fluctuating \textit{spin fields}. Although the spin fields arise from the electrons microscopically, we treat them as independent degrees of freedom in the spirit of the spin-fermion model \cite{Abanov2003quantum}. In this work, we consider a spin-fermion model of Dirac fermions which can be realized on a honeycomb lattice, $\hat{H}=\hat{H}_e+\hat{H}_s+\hat{H}_{es}$.
The low-energy electron is described by \cite{Wu2019_toplogical}
\begin{align}\label{Eq:H_e}
	\hat{H}_e=\sum_{\vex{k}}\psi^{\dagger}_{\vex{k}}h_{\vex{k}}\psi_{\vex{k}},
\end{align}
where 
\begin{align}\label{Eq:h_k}
h_{\vex{k}}= \hbar v_F\sigma^x\tau^zk_x+ \hbar v_F\sigma^yk_y-E_F.
\end{align}
In the above expression, $\sigma^a$ ($\tau^a$) denotes the $a$-component of the Pauli matrices for the sublattice (valley), $v_F$ is the Dirac velocity, $E_F$ is the Fermi energy, and $\psi_{\vex{k}}$ is an 8-component fermionic field (valley, sublattice, and the \textit{intrinsic} spin) with momentum $\vex{k}$. The reduced Hamiltonian $h_{\vex{k}}$ obeys a ($\mathcal{T}^2=1$) time-reversal operator, $\hat{\Pi}h^*_{-\vex{k}}\hat{\Pi}^{-1}=h_{\vex{k}}$, where $\hat{\Pi}=\tau^x$. The fluctuating spin fields are described by 
\begin{align}\label{Eq:H_s}
	\hat{H}_s=&\frac{1}{2}\sum_{\vex{q}}\chi^{-1}(\vex{q})\vec{S}_{-\vex{q}}\cdot\vec{S}_{\vex{q}},
\end{align}
where $\chi^{-1}(\vex{q})$ is the inverse spin-spin zero-frequency correlation function and $\vec{S}_{\vex{q}}=(S^x_{\vex{q}},S^y_{\vex{q}},S^z_{\vex{q}})$ encodes the three-component fluctuating bosonic spin field at momentum $\vex{q}$. Finally, the spin-fermion coupling is given by
\begin{align}\label{Eq:H_es}
\hat{H}_{es}=\frac{g}{\sqrt{A}}\sum_{\vex{k},\vex{q}}\left(\psi^{\dagger}_{\vex{k}-\vex{q}}\,\frac{\vec{\mu}}{2}\,\psi_{\vex{k}}\right)\cdot\vec{S}_{\vex{q}},
\end{align}
where $g$ is the spin-fermion coupling constant, $A$ is the area of the system, and the Pauli matrices for the spin $\vec{\mu}=(\mu^x,\mu^y,\mu^z)$. The spin-fermion model here is motivated by the non-spin-singlet superconductivity in the MATTG experiment near $\nu=-2$ \cite{Cao2021} since spin fluctuation provides a natural explanation for the spin triplet pairing \cite{Coleman2015introduction,Mineev2017superconductivity}.
We further assume that the fluctuating spin fields fail to develop a long range order at the temperatures of our interest, but the spin-spin correlation function $\chi(\vex{q})$ is peaked at $\vex{q}=0$, e.g., $\chi(\vex{q})=\chi_0/(|\vex{q}|^2+\xi^{-2})$ where $\chi_0>0$ and $\xi$ is the correlation length. 
Such an assumption is consistent with the absence of ground state magnetization in MATTG experiment near $\nu=-2$ \cite{Cao2021}.

To obtain an effective inter-electron interaction, we integrate out the fluctuating spin fields in Eqs.~(\ref{Eq:H_s}) and (\ref{Eq:H_es}). The effective interaction is given by
\begin{align}
\label{Eq:H_I}\hat{H}_I\!=\!-\frac{g^2}{2A}\sum_{\vex{k},\vex{k}',\vex{q}}\chi(\vex{q})\left(\!\psi^{\dagger}_{\vex{k}+\vex{q}}\,\frac{\vec{\mu}}{2}\,\psi_{\vex{k}}\!\right)\!\cdot\!\left(\!\psi^{\dagger}_{\vex{k}'-\vex{q}}\,\frac{\vec{\mu}}{2}\,\psi_{\vex{k}'}\!\right).
\end{align}
$\hat{H}_I$ describes the ferromagnetic interaction between the intrinsic spins of the itinerant electrons which favors spin-triplet pairing \cite{Coleman2015introduction,Mineev2017superconductivity,SM}. One possibility is paramagnon mediated interactions.

\textit{Pairing symmetry.--} Besides the spin degrees of freedom, the valley and sublattice structures play important roles in the pairing symmetry of the superconductivity \cite{Wu2018,Wu2019_phonon}. We discuss only the inter-valley Cooper pairs as the intra-valley Cooper pairs correspond to a lattice-scale oscillating gap function in the position space. The $s$-wave and $f$-wave pairings are intra-sublattice while the $p$-wave and $d$-wave pairings are inter-sublattice \cite{Wu2019_phonon}. 
To see this, we consider Cooper pairs in the position space and examine the symmetry operations in the following. 
The wavefunction under the three-fold rotation ($\mathcal{C}_{3z}$) about a hexagon center is given by $\mathcal{C}_{3z}\psi^{\dagger}(\vex{r})\mathcal{C}_{3z}^{-1}=e^{i\frac{2\pi}{3}\tau_z\sigma_z}\psi^{\dagger}(\mathcal{R}_3\vex{r})$ as illustrated in Fig.~\ref{Fig:Sub_pairing}(b). In addition, the two-fold rotation ($\mathcal{C}_{2z}$) about a hexagon center is implemented by the valley and sublattice exchanging as follows: $\mathcal{C}_{2z}\psi^{\dagger}(\vex{r})\mathcal{C}_{2z}^{-1}=\tau^x\sigma^x\psi^{\dagger}(\mathcal{R}_2\vex{r})$.
For a pair of electrons from different valleys, $\mathcal{C}_{3z}$ operation can distinguish the $s$-wave and $f$-wave ($L_z$ mod $3=0$) from the $p$-wave and $d$-wave ($L_z$ mod $3=1,2$); $\mathcal{C}_{2z}$ operation can distinguish the $s$-wave and $d$-wave ($L_z$ mod $2=0$) from the $p$-wave and $f$-wave ($L_z$ mod $2=1$). 
With both $\mathcal{C}_{3z}$ and $\mathcal{C}_{2z}$, we can classify angular momentum states associated with $|L_z|=0,1,2,3$.
Therefore, the sublattice structures are fully determined for the inter-valley $s$-, $p$-, $d$-, and $f$-wave pairings \cite{Wu2018,Wu2019_phonon}. 

The spin-triplet Cooper pair bilinears are summarized in the following. (The spin-singlet operators are listed in Sec. I of \cite{SM}.) The $p$-wave spin-triplet Cooper pairs are inter-sublattice and are given by
\begin{align}
	\vec{\Phi}_{p,X}(\vex{k})=&\psi^T_{-\vex{k}}\left[(-i\tau^y)\sigma^x\left(i\vec{\mu}\mu^y\right)^{\dagger}\right]\psi_{\vex{k}},\\
	\vec{\Phi}_{p,Y}(\vex{k})=&\psi^T_{-\vex{k}}\left[\tau^x(i\sigma^y)\left(i\vec{\mu}\mu^y\right)^{\dagger}\right]\psi_{\vex{k}}.
\end{align}
The subscripts $X$ and $Y$ correspond to the $p_x$ and $p_y$ structure respectively. The $f$-wave spin-triplet Cooper pairs are described by
\begin{align}\label{Eq:CP_f}
	\vec{\Phi}_{f}(\vex{k})=&\psi^T_{-\vex{k}}\left[(-i\tau^y)\left(i\vec{\mu}\mu^y\right)^{\dagger}\right]\psi_{\vex{k}}.
\end{align}
In addition to the $p$-wave and $f$-wave, we find a \textit{staggered} intra-sublattice spin-triplet Cooper pair \cite{Staggered} given by
\begin{align}\label{Eq:CP_S}
	\vec{\Phi}_{\tilde{s}}(\vex{k})=&\psi^T_{-\vex{k}}\left[(-i\tau^y)\sigma^z\left(i\vec{\mu}\mu^y\right)^{\dagger}\right]\psi_{\vex{k}}.
\end{align}
The $\sigma^z$ indicates a staggered sublattice structure, which might suggest a vanishingly small effect, similar to the intra-valley pairing terms. 

The above bilinear exhaust all the possible spin-triplet pairing.
We note that all the spin-triplet Cooper pairs discussed above can have a pairing potential with no explicit dependence on momentum $\vex{k}$. This should be contrasted with other spin-triplet superconductors such as $^3$He \cite{Coleman2015introduction} and heavy fermion ferromagnetic superconductors \cite{Ran2019nearly,Mineev2017superconductivity}, where the spin-triplet pairing potential is an odd function of momentum $\vex{k}$. 

Notice that it is important to use the bilinear operators with respect to the basis in Eq.~(\ref{Eq:H_e}) because the expressions of Cooper pairs are basis-dependent.  We adopt the microscopic basis for Eq.~(\ref{Eq:H_e}) and the bilinear operators discussed above.

\textit{BCS theory.--} To investigate the qualitative features of the superconductivity, we employ the standard BCS approach \cite{Coleman2015introduction,Samokhin2008} and derive the pairing Hamiltonian from Eq.~(\ref{Eq:H_I}) (with the attractive channels only) as follows:
\begin{align}
	\nonumber\hat{H}_{I}'=&-\frac{\tilde{U}}{A}\sum'_{\vex{k},\vex{k}'}\bigg[\vec{\Phi}^{\dagger}_{p,X}(\vex{k})\cdot\vec{\Phi}_{p,X}(\vex{k}')+\vec{\Phi}^{\dagger}_{p,Y}(\vex{k})\cdot\vec{\Phi}_{p,Y}(\vex{k}')
	\\
	\label{Eq:pairing_H}&+\vec{\Phi}^{\dagger}_{\tilde{s}}(\vex{k})\cdot\vec{\Phi}_{\tilde{s}}(\vex{k}')+\vec{\Phi}^{\dagger}_{f}(\vex{k})\cdot\vec{\Phi}_{f}(\vex{k}')\bigg],
\end{align}
where $\tilde{U}>0$ is the momentum-independent effective pairing strength, the prime in the momentum summation indicates summing half of the Brillouin zone. The interaction $\tilde{U}$ is momentum-independent, but we still find unconventional pairings such as $p$-wave and $f$-wave because of the valley and sublattice degrees of freedom.
$\tilde{U}$ can be derived from $\chi(\vex{q})$ in Eq.~(\ref{Eq:H_I}), which is discussed in Sec. II of \cite{SM}.
Equation (\ref{Eq:pairing_H}) describes attractive interactions among the spin-triplet Cooper pairs, implying spin-triplet superconductivity as the leading instability. The spin-singlet terms, which are left out in Eq.~(\ref{Eq:pairing_H}) are given in Sec. II of \cite{SM} and are checked to be repulsive as expected from ferromagnetic spin-fluctuation pairing \cite{Coleman2015introduction}.

To study the superconductivity, we employ the mean-field decoupling in Eq.~(\ref{Eq:pairing_H}) and express the mean field theory in terms of the Bogoliubov-de Gennes (BdG) Hamiltonian as follows:
\begin{align}
	\nonumber\hat{H}_{\text{MFT}}\!=&\sum_{\vex{k}}'\Psi_{\vex{k}}^{\dagger}\mathcal{H}_{\text{BdG}}(\vex{k})\Psi_{\vex{k}}\\
	\label{Eq:H_BdG}	&+\!\frac{A}{\tilde{U}}\!\left(\!\left|\vec{\Delta}_{p,X}\right|^2\!+\!\left|\vec{\Delta}_{p,Y}\right|^2\!+\!\left|\vec{\Delta}_{f}\right|^2\!+\!\left|\vec{\Delta}_{\tilde{s}}\right|^2\right),
\end{align}
where $\Psi_{\vex{k}}^T=[\psi_{\vex{k}}^T,\,\psi^{\dagger}_{-\vex{k}}(-i\mu^y)]$, $\vec{\Delta}_{p,X}$ ($\vec{\Delta}_{p,Y}$) is the order parameters for the $p_x$ ($p_y$) pairing, and $\vec{\Delta}_f$ ($\vec{\Delta}_{\tilde{s}}$) is the order parameter for the $f$-wave (staggered intra-sublattice) pairing. $\mathcal{H}_{\text{BdG}}(\vex{k})$ is expressed by
\begin{align}
	\nonumber\mathcal{H}_{\text{BdG}}(\vex{k})=&(\hbar v_Fk_x)\sigma^x\tau^z\hat{1}_{\kappa}+(\hbar v_Fk_y)\sigma^y\kappa^z-E_F\kappa^z\\
	&+\left(\vec{\Xi}\cdot\vec{\mu}\right)\kappa^++\left(\vec{\Xi}^{\dagger}\cdot\vec{\mu}\right)\kappa^-,\\
	\nonumber\vec{\Xi}=&\vec{\Delta}_{p,X}(-i\tau^y)\sigma^x+\vec{\Delta}_{p,Y}\tau^x(i\sigma^y)\\
	&+\vec{\Delta}_{f}(-i\tau^y)+\vec{\Delta}_{\tilde{s}}(-i\tau^y)\sigma^z,
\end{align}
where the Pauli matrices ($\kappa^{x,y,z}$) and identity ($\hat{1}_{\kappa}$) in the particle-hole space are introduced and $\kappa^{\pm}=(\kappa^x\pm i\kappa^y)/2$.  One can easily confirm the particle-hole symmetry
\begin{align}
	\hat{P}\mathcal{H}^*_{\text{BdG}}(-\vex{k})\hat{P}^{-1}=-\mathcal{H}_{\text{BdG}}(\vex{k}),
\end{align}
where $\hat{P}=\mu^y\kappa^y$, corresponding to the $\mathcal{P}^2=1$ particle-hole symmetry.

\begin{figure}[t!]
	\includegraphics[width=0.35\textwidth]{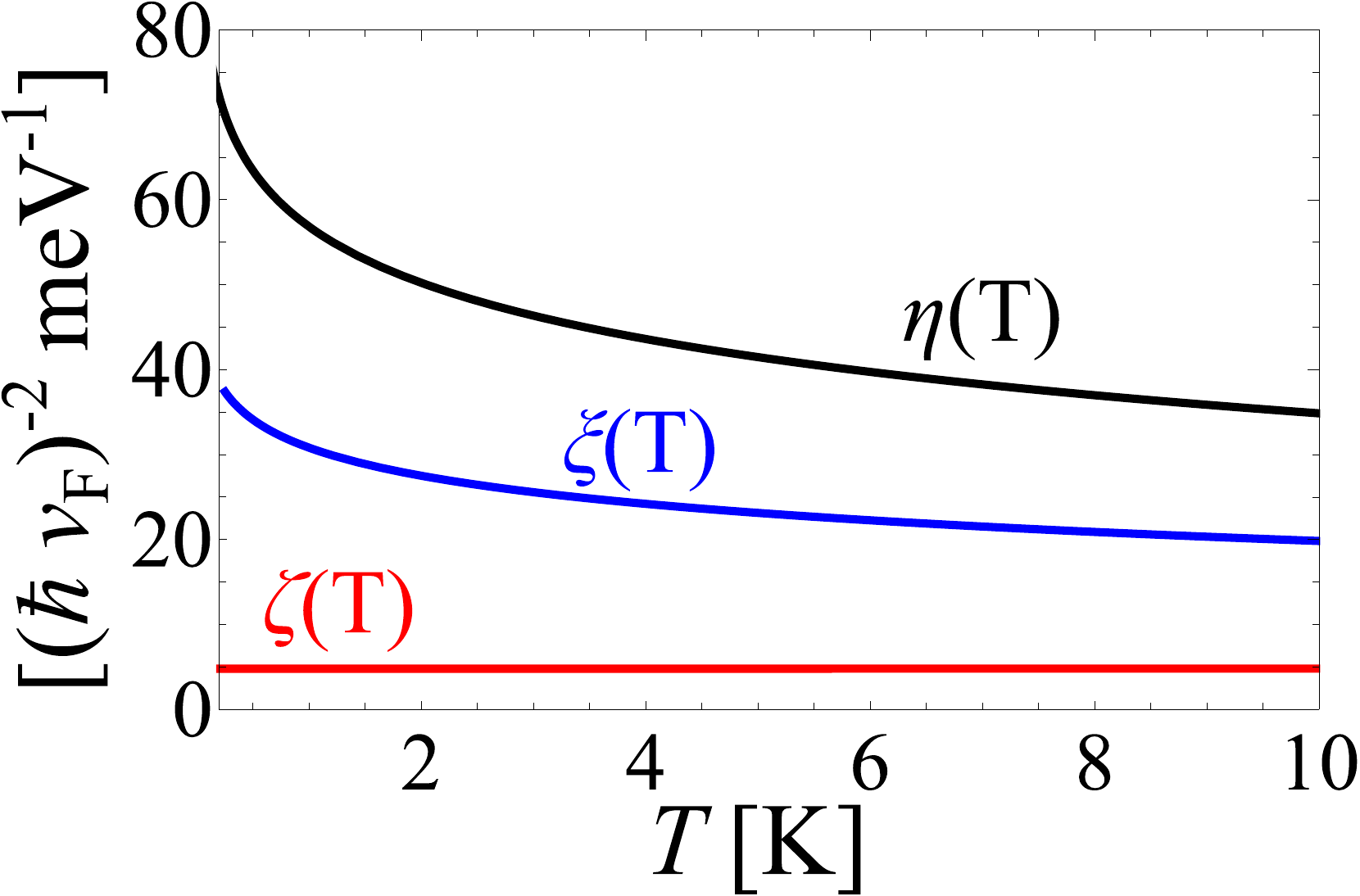}
	\caption{Coefficients in the Landau free energy density. We plot $\eta(T)$ (black curve), $\xi(T)$ (blue curve), and $\zeta(T)$ (red curve) with $E_F=30$meV and $\hbar v_F\Lambda=50$meV. The unit in the $y$-axis is $\left(\hbar v_F\right)^{-2}$meV$^{-1}$. The results show that $\tilde{\eta}(T)>\tilde{\xi}(T)>\tilde{\zeta}(T)$ in the parameter regime relevant to MATTG experiment, suggesting that the $f$-wave is the dominating pairing symmetry. }
	\label{Fig:Coeff}
\end{figure}

Formally, we can derive the free energy associated with Eq.~(\ref{Eq:H_BdG}) by integrating out the fermionic field in the partition function. Then, the free energy density ($\mathcal{F}$) is expressed by
\begin{align}
\nonumber\mathcal{F}=&-\frac{k_BT}{A}\ln\det\left[-i\omega_n\hat{1}+\mathcal{H}_{\text{BdG}}(\vex{k})\right]\\
\label{Eq:Free_f_form}&+\frac{1}{\tilde{U}}\left(\left|\vec{\Delta}_{\tilde{s}}\right|^2+\left|\vec{\Delta}_{p,X}\right|^2+\left|\vec{\Delta}_{p,Y}\right|^2+\left|\vec{\Delta}_{f}\right|^2\right),
\end{align}
where $k_B$ is the Boltzmann constant.
We perform a perturbative expansion of $\mathcal{F}$ upto the quadratic order of the order parameter. This can be done systematically with the standard treatment in the matrix Green function approach \cite{Coleman2015introduction}. The Landau-type free energy density is given by
\begin{align}
	\nonumber \mathcal{F}\!=\!&\text{ const}+\left[\frac{1}{\tilde{U}}-\xi(T)\right]\left(|\vec{\Delta}_{p,X}|^2+|\vec{\Delta}_{p,Y}|^2\right)\\
	\label{Eq:FreeF_exp}&\!+\!\left[\frac{1}{\tilde{U}}\!-\!\eta(T)\!\right]\!\!|\vec{\Delta}_{f}|^2\!+\!\!\left[\frac{1}{\tilde{U}}\!-\!\zeta(T)\!\right]\!\!|\vec{\Delta}_{\tilde{s}}|^2\!+\!O(|\Delta|^4),
\end{align}
where $\eta(T)>\xi(T)>\zeta(T)>0$ (see Fig.~\ref{Fig:Coeff} and Sec.~IV in \cite{SM}), and $T$ is the temperature. The transition temperatures can be determined by $\tilde{U}\xi(T_c^p)=1$, $\tilde{U}\eta(T_c^f)=1$, and $\tilde{U}\zeta(T_c^{\tilde{s}})=1$, where $T_c^p$, $T_c^f$, and $T_c^{\tilde{s}}$ are the transition temperatures for the $p$-wave, $f$-wave, and the staggered intra-sublattice spin-triplet pairings respectively. 
Remarkably, we obtain that $T_c^f>T_c^p>T_c^{\tilde{s}}$, suggesting that the $f$-wave is the dominating superconducting state. 
This result is due to the valley-sublattice structure \cite{Wu2018,Wu2019_phonon} (see Sec.~IV in \cite{SM} for a discussion) rather than the detail band structure, and we anticipate that $f$-wave is generically the dominating pairing symmetry for spin-triplet superconductivity in the graphene based systems including MATTG, independent of the pairing mechanism.
We note that the possibility of realizing $f$-wave superconductivity was also discussed previously in the context of graphene and MATBG \cite{Nandkishore2014,Lin2018,Wu2020,Wang2021,Wu2019_phonon,Throckmorton2020,Alidoust2019,Szabo2021}.

\textit{$f$-wave superconductivity.--} In the rest of this Letter, we focus on the $f$-wave pairing spin-triplet superconductivity.  With only the $f$-wave order parameter, the free energy density given by Eq.~(\ref{Eq:Free_f_form}) can be derived exactly (see Sec. V of \cite{SM}) and is expressed by
\begin{align}
\label{Eq:FreeE_f}\mathcal{F}=&-2k_BT\sum_{s,r=\pm}\int_{\vex{k}}\ln\left[2\cosh\left(\frac{\sqrt{\epsilon_{\vex{k},s}^2+\Delta_r^2}}{2k_BT}\right)\right]+\frac{|\vec{\Delta}_f|^2}{\tilde{U}},
\end{align}
where $\epsilon_{\vex{k}.\pm}=\hbar v_F|\vex{k}|\pm E_F$, $\Delta_{\pm}=\sqrt{|\vec{\Delta}_f|^2\pm|i\vec{\Delta}_f\times\vec{\Delta}_f^*|}$, and $\int_{\vex{k}}$ denotes $\int\frac{d^2\vex{k}}{(2\pi)^2}$. Without a magnetic field, the free energy is minimized when $i\vec{\Delta}_f\times\vec{\Delta}_f^*=0$ (i.e., $\vec{\Delta}_f\parallel\vec{\Delta}_f^*$). Therefore, the order parameter can be expressed by $\vec{\Delta}_f=e^{i\phi}\vec{O}$, where $\vec{O}$ is a real-valued vector.
Note that the phase $\phi$ can be gauged away. We find that the $f$-wave BdG Hamiltonian satisfies a ($\mathcal{T}^2=1$) time-reversal symmetry:
\begin{align}\label{Eq:TRO}
	\hat{\Pi}\mathcal{H}^*_{\text{BdG}}(-\vex{k})\hat{\Pi}^{-1}=\mathcal{H}_{\text{BdG}}(\vex{k}),
\end{align}
where $\hat{\Pi}=\sigma^y\tau^z\mu^y$ \cite{TR}.
The superconducting state here belongs to the class BDI, which is topologically trivial in two dimensions based on the ten-fold way classification \cite{Schnyder2008}.
As such, we conclude that the spin-triplet $f$-wave pairing superconducting state is \textit{unitary} (i.e., $i\vec{\Delta}_f\times\vec{\Delta}_f^*=0$), time-reversal symmetric, fully gapped, and topologically trivial. 

To derive the gap equation, we minimize the free energy in Eq.~(\ref{Eq:FreeE_f}) with respect to $|\vec{\Delta}_f|^2$ (with $i\vec{\Delta}_f\times\vec{\Delta}_f^*=0$). The gap equation is expressed by
\begin{align}
\frac{1}{\tilde{U}}=\int_{\vex{k}}\left[\frac{\tanh\left(\frac{E_{\vex{k},+}}{2k_BT}\right)}{E_{\vex{k},+}}+\frac{\tanh\left(\frac{E_{\vex{k},-}}{2k_BT}\right)}{E_{\vex{k},-}}\right],
\end{align}
where $E_{\vex{k},\pm}=\sqrt{\epsilon_{\vex{k},\pm}^2+|\vec{\Delta}_f|^2}$. To regularize the momentum space integral, we introduce a momentum cutoff ($\Lambda$). For $|\vec{\Delta}_f|\ll |E_F|, \hbar v_F\Lambda$, we derive the asymptotic expressions for the zero temperature gap, $2\Delta_0\equiv2|\vec{\Delta}_f(T=0)|$, and the transition temperature, $T_c^f$, as follows \cite{DP}:
\begin{align}
\label{Eq:Delta_0}2\Delta_0=&2|E_F|\mathcal{A}\exp\left[-\frac{1}{\tilde{U}N(E_F)}\right],\\
\label{Eq:T_c}k_BT_c^f=&|E_F| \frac{e^{\gamma}}{\pi}\mathcal{A}\exp\left[-\frac{1}{\tilde{U}N(E_F)}\right],
\end{align}
where $\mathcal{A}=2\sqrt{\frac{\hbar v_F\Lambda-|E_F|}{\hbar v_F\Lambda+|E_F|}}\exp\left(\frac{\hbar v_F\Lambda-|E_F|}{|E_F|}\right)$ is a dimensionless parameter, $\gamma$ is the Euler–Mascheroni constant, and $N(E_F)=|E_F|/(2\pi \hbar^2v_F^2)$ is the density of states at the Fermi energy.

\begin{figure}[t!]
	\includegraphics[width=0.25\textwidth]{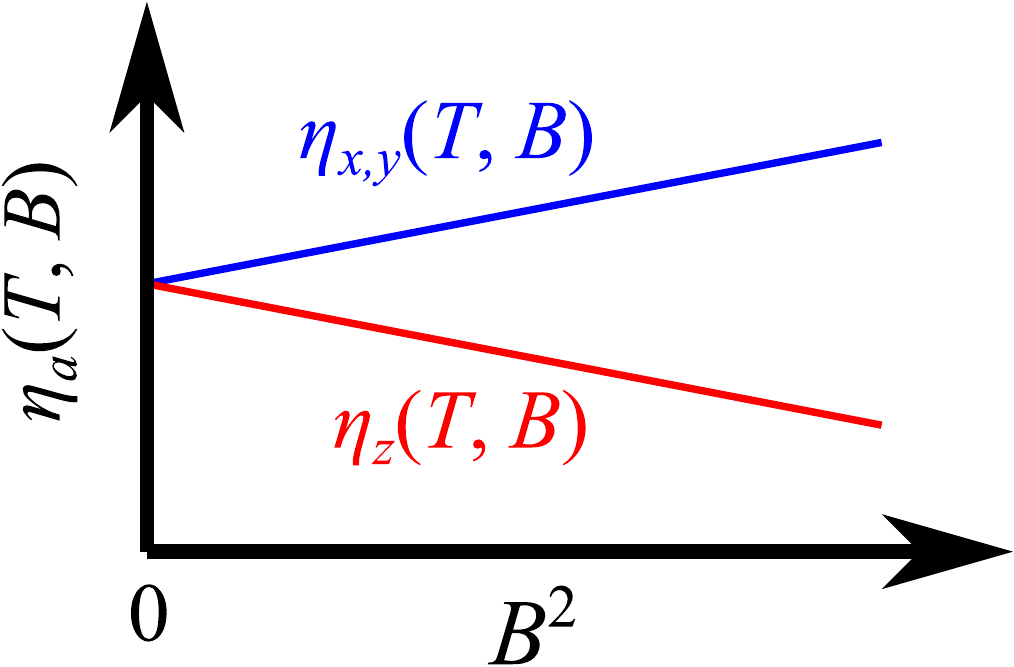}
	\caption{Sketch of $\eta_a(T,B)$ versus $B^2$. In the absence of Zeeman splitting (i.e., $B=0$), $\eta_x(T,0)=\eta_y(T,0)=\eta_z(T,0)$. For a small finite $B$, $\eta_x(T,B)=\eta_y(T,B)>\eta_z(T,B)$ and $\eta_x(T,B)-\eta_z(T,B)\propto B^2$. See main text for a detailed discussion.}
	\label{Fig:B_field}
\end{figure}

To gain an intuitive understanding, we expand the free energy density (without an applied magnetic field) in Eq.~(\ref{Eq:FreeE_f}) upto the quartic order of the order parameter as follows:
\begin{align}
\nonumber \mathcal{F}=&\text{ const}+\sum_{a=x,y,z}\left[\frac{2}{g^2U}-\eta_a(T)\right]|\Delta_f^a|^2\\
\label{Eq:FreeE_f_exp}&+a_4(T)\left[|\vec{\Delta}_f|^4+|i\vec{\Delta}_f^*\times\vec{\Delta}_f|^2\right]+O(|\Delta_f|^6),
\end{align}
where $a_4(T)>0$ and $\eta_a(T)=\eta(T)$ in the absence of a magnetic field. The absence of $i\vec{\Delta}_f\times\vec{\Delta}^*_f$ is the manifestation of the positive $a_4(T)$ in the Landau theory. 

In the presence of a small in-plane magnetic field, electronic band develops a Zeeman splitting. Then, $-B\mu^z$ is added to $h_{\vex{k}}$ [Eq.~(\ref{Eq:h_k})], where $2B$ denotes the Zeeman splitting and the $z$ direction is rotated to the field direction.
Similar to the expansion in Eq.~(\ref{Eq:FreeF_exp}), we treat the order parameter perturbatively. In addition, we also treat $-B\mu^z$ perturbatively in the Green functions. (See Sec.~VI of \cite{SM} for a derivation.)
The free energy density acquires a term proportional to $-iB\vec{\Delta}_f\times\vec{\Delta}^*_f\cdot\hat{z}$, implying  $i\vec{\Delta}_f\times\vec{\Delta}^*_f\cdot\hat{z}\neq 0$ and a finite magnetization in the superconducting state. The quadratic terms in the free energy density are also affected by $B$ as illustrated in Fig.~\ref{Fig:B_field}. Specifically, $\eta_x(T,B)=\eta_y(T,B)=\eta(T,0)+B^2\delta\eta(T)$ and $\eta_z(T,B)=\eta(T,0)-B^2\delta\eta(T)$, where $\delta\eta(T)>0$ (Sec. VI of \cite{SM}). Thus, $\Delta_f^x$ and $\Delta_f^y$ are the favored states, suggesting that equal-spin pairing is the dominating pairing structure with a Zeeman splitting [See Eq.~(\ref{Eq:CP_f})]. Recall that $\Delta_{\pm}=\sqrt{|\vec{\Delta}_f|^2\pm |i\vec{\Delta}_f\times\vec{\Delta}^*_f|}$. Thus, there are two gaps in the superconducting state. For a small $B$, the gap formed by `up' spins is increased while the gap formed by the `down' spins is suppressed. This is consistent with the previous theoretical results in \cite{Wu2019_identification} for MATBG. We conclude that the superconductivity with a small in-plane field is $f$-wave, spin-triplet, and equal-spin pairing.

\textit{Relation to MATTG experiment.--} A recent MATTG experiment found that the superconductivity near $\nu=-2$ persists for a large magnetic field (10T) beyond the Pauli limit for the spin-singlet superconductivity \cite{Cao2021}. A natural explanation is that the superconductivity is spin-triplet, which we emphasize in this Letter. However, there is no clear sign of magnetism in the experiments. A possible explanation is that the electrons are spin polarized, but the coherence length is too small to develop a true long-range order. In such a scenario, the phenomenological spin-fermion model provides a natural starting point. In addition, our predicted $f$-wave superconducting state can withstand finite Zeeman splitting, and the gap (from the `up' spins) is enhanced for a small in-plane magnetic field. To compare with the MATTG experiments, we take $v_F=4\times 10^4$m/s, $\hbar v_F\Lambda=50$meV (half of the MATTG bandwidth \cite{Park2021tunable}), $|E_F|=30$meV, and $\tilde{U}=30$meV$\cdot$nm$^2$. Putting these numbers in Eq.~(\ref{Eq:T_c}), we obtain $T_c^f=3$K, which is comparable to the extracted $T_{\text{BKT}}$ in the experiments \cite{Park2021tunable,Hao2021electric}. We also note that $\tilde{U}$ cannot be larger than $\tilde{U}_c\sim 1/\rho_0$ (with $\rho_0$ being the density of states at Fermi level) as the Stoner ferromagnetism must be absent in the normal state. Although the spin-fluctuation mechanism proposed in this work is likely, we cannot not rule out the acoustic-phonon mechanism which might realize $f$-wave spin-triplet superconductivity also \cite{Wu2019_phonon,Wu2020_TDBG,Chou2021SC_RTG}.

The $f$-wave spin-triplet superconductivity can be examined experimentally. The spin-triplet Cooper pair cannot tunnel into a spin-singlet superconductor (i.e., zero Josephson current) but can tunnel into an ``Ising'' superconductor (such as NbSe$_2$ \cite{Xi2016ising,Kim2017strong}) as long as the spin-triplet Cooper pair has zero out-of-plane spin projection. Since the small magnetic field can partially polarize the superconducting state, the MATTG-NbSe$_2$ junction with a small magnetic field can distinguish the spin-triplet pairing from the spin-singlet pairing.
In addition, the $f$-wave symmetry can be confirmed by a hybrid corner Josephson junction as discussed in Sec. VII of \cite{SM}.

\textit{Relation to Hubbard model.--} The phenomenological spin-fermion model used in this work might be derived from the Hubbard model \cite{Abanov2003quantum}. In the single-band square lattice $SU(2)$ Hubbard model, the antiferromagnetism arises at half filling for a large onsite repulsion, and the Nagaoka ferromagnetism takes place when doping slightly away from the half filling. We expect that the phenomenological spin-fermion model here should capture the gross features of the possible spin polarized states in the MATTG bands. However, the microscopic justification remains an important question, both whether there is superconductivity in the Hubbard model \cite{Arovas2021hubbard} and whether Hubbard model is relevant to MATTG \cite{Fischer2021unconventional,Roy2019,Szabo2021}. Our work, however, transcends these questions and applies as long as spin fluctuations mediate the observed superconductivity.

\textit{Outlook.--} The present work can be straightforwardly generalized to the MATTG bands \cite{Li2019,Khalaf2019,Cualuguaru2021}, which is an important direction for future work. We anticipate that $f$-wave remains the main pairing symmetry because of the valley-sublattice structure unique to graphene. In this Letter, we concentrate only on the zero and the low in-plane magnetic field limits. With a sufficiently large in-plane magnetic field, the spin fields are fully polarized, and a sizable Zeeman splitting in the electronic bands develop. The transverse fluctuation of the polarized spin fields (analogous to the magnon) can still mediate the effective interaction. The presence of spin fluctuations implies that the Pomenranchauk effect applies to the system \cite{Rozen2021entropic,Saito2021isospin}.

\begin{acknowledgments}
We are grateful to Andrey Chubukov, Pablo Jarillo-Herrero, and Andrea Young for stimulating correspondence.
Y.-Z.C. also thanks Zhentao Wang for useful discussions.
This work is supported by the Laboratory for Physical Sciences (Y.-Z.C. and S.D.S.), by JQI-NSF-PFC (supported by NSF
Grant No. PHY-1607611, Y.-Z.C.), and NSF DMR1555135 (CAREER, J.D.S.). F.W. is supported by startup funding of Wuhan University.
\end{acknowledgments}	
	




\newpage \clearpage 

\onecolumngrid

\begin{center}
	{\large
		Correlation-induced triplet pairing superconductivity in graphene-based moir\'e systems
		\vspace{4pt}
		\\
		SUPPLEMENTAL MATERIAL
	}
\end{center}

\setcounter{figure}{0}
\renewcommand{\thefigure}{S\arabic{figure}}
\setcounter{equation}{0}
\renewcommand{\theequation}{S\arabic{equation}}

In this supplemental material, we provide some technical details for main results in the main text. We set $\hbar=1$ and $k_B=1$ in this supplemental material.

\section{Spin-singlet Cooper pair bilinears}

We provide a list of spin-singlet Cooper pair operators as follows:
\begin{align}
	\Phi_{s}(\vex{k})=&\psi^T_{-\vex{k}}\left[\tau^x(i\mu^y)^{\dagger}\right]\psi_{\vex{k}},\\
	\Phi_{d,S}(\vex{k})=&\psi^T_{-\vex{k}}\left[\tau^x\sigma^x\left(i\mu^y\right)^{\dagger}\right]\psi_{\vex{k}},\\
	\Phi_{d,A}(\vex{k})=&\psi^T_{-\vex{k}}\left[(-i\tau^y)(i\sigma^y)\left(i\mu^y\right)^{\dagger}\right]\psi_{\vex{k}},\\
	\Phi_{\tilde{s},0}(\vex{k})=&\psi^T_{-\vex{k}}\left[\tau^x\sigma^z(i\mu^y)^{\dagger}\right]\psi_{\vex{k}},
\end{align}
where $\Phi_{s}$ is the $s$-wave spin-singlet Cooper pair $\Phi_{d,S}$ ($\Phi_{d,A}$) is the $d$-wave spin-singlet Cooper pair with symmetric (antisymmetric) sublattice structure, $\Phi_{\tilde{s},0}$ is the staggered intra-sublattice spin-singlet Cooper pair. Physically, $\Phi_{\tilde{s},0}$ is unlikely to be important because of the staggered sublattice structure, similar to the $\vec{\Phi}_{\tilde{s}}$ discussed in the main text.

\section{Pairing Hamiltonian}

The pairing Hamiltonian including both the spin-singlet and spin-triplet channels is expressed as follows:
\begin{align}
	\nonumber\hat{H}_{I}'
	=&-\frac{\tilde{U}}{A}\sum'_{\vex{k},\vex{k}'}\left[\sum_{a=X,Y}\vec{\Phi}^{\dagger}_{p,Y}(\vex{k})\cdot\vec{\Phi}_{p,Y}(\vex{k}')+\vec{\Phi}^{\dagger}_{f}(\vex{k})\cdot\vec{\Phi}_{f}(\vex{k}')+\vec{\Phi}^{\dagger}_{s}(\vex{k})\cdot\vec{\Phi}_{s}(\vex{k}')\right]\\
	&+\frac{3\tilde{U}}{A}\sum'_{\vex{k},\vex{k}'}\left[\Phi^{\dagger}_{s0}(\vex{k})\Phi_{s0}(\vex{k}')+\sum_{a=S,A}\Phi^{\dagger}_{d,a}(\vex{k})\Phi_{d,a}(\vex{k}')+\Phi^{\dagger}_{\tilde{s}0}(\vex{k})\Phi_{\tilde{s}0}(\vex{k}')\right].
\end{align}
The above equation describes attractive interactions between spin-triplet pairs and repulsive interactions between spin-singlet pairs. Thus, the spin-singlet superconductivity is not possible with the ferromagnetic interaction as expected.

It is important to emphasize that the spin-triplet channel in $\hat{H}_I'$ is parity-even rather than parity-odd. Such an unexpected property is allowed by the valley-sublattice structure in the honeycomb lattice.  
Microscopically, $\tilde{U}=g^2\tilde{\chi}/4$, where
\begin{align}
	\tilde{\chi}=\frac{1}{2}\int\frac{d\theta}{2\pi}\left[\chi(\vex{k}+\vex{k}')+\chi(\vex{k}-\vex{k}')\right]\bigg|_{|\vex{k}|=|\vex{k}'|=k_F},
\end{align}
and $\theta$ is the angle between $\vex{k}$ and $\vex{k}'$.
$\tilde{\chi}$ is the angular average of the symmetrized correlation function evaluated at the Fermi surface, which does not depend on $\vex{k}$ or $\vex{k}'$. The parity-even property of the interaction is due to the antisymmetrized spin-triplet Cooper pairs allowed by valley and sublattice. 

\section{Mean field decoupling}

To study the superconductivity, we employ the mean-field decoupling and obtain the mean-field Hamiltonian as follows:
\begin{align}
	\nonumber\hat{H}_{\text{MFT}}=&\sum_{\vex{k}}\psi^{\dagger}_{\vex{k}}h_{\vex{k}}\psi_{\vex{k}}+\sum_{\vex{k}}'\bigg[\vec{\Phi}^{\dagger}_{\tilde{s}}(\vex{k})\cdot\vec{\Delta}_{\tilde{s}}+\vec{\Phi}^{\dagger}_{f}(\vex{k})\cdot\vec{\Delta}_{f}+\vec{\Phi}^{\dagger}_{p,X}(\vex{k})\cdot\vec{\Delta}_{p,X}+\vec{\Phi}^{\dagger}_{p,Y}(\vex{k})\cdot\vec{\Delta}_{p,Y}+\text{H.c.}\bigg]\\
	\label{Eq:MFT}&+\frac{A}{\tilde{U}}\left(\left|\vec{\Delta}_{p,X}\right|^2+\left|\vec{\Delta}_{p,Y}\right|^2+\left|\vec{\Delta}_{f}\right|^2+\left|\vec{\Delta}_{\tilde{s}}\right|^2\right),
\end{align}
where $\vec{\Delta}_{p,X}$ ($\vec{\Delta}_{p,Y}$) is the order parameters for the $p_x$ ($p_y$) pairing and  $\vec{\Delta}_f$ ($\vec{\Delta}_{\tilde{s}}$) is the order parameter for the $f$-wave (staggered intra-sublattice) pairing. The mean field Hamiltonian can be recast to a compact Bogoliubov-de Gennes (BdG) form as follows:
\begin{align}
	\hat{H}_{MFT}=&\sum_{\vex{k}}'\Psi_{\vex{k}}^{\dagger}\mathcal{H}_{\text{BdG}}(\vex{k})\Psi_{\vex{k}}+\frac{A}{\tilde{U}}\left(\left|\vec{\Delta}_{p,X}\right|^2+\left|\vec{\Delta}_{p,Y}\right|^2+\left|\vec{\Delta}_{f}\right|^2+\left|\vec{\Delta}_{\tilde{s}}\right|^2\right),
\end{align}
where
\begin{align}
	\Psi_{\vex{k}}=&\left[\begin{array}{c}
		\psi_{\vex{k}}\\[1mm]
		i\mu^y\left(\psi^{\dagger}_{-\vex{k}}\right)^T
	\end{array}\right],\\
	\mathcal{H}_{\text{BdG}}(\vex{k})=&\left[\begin{array}{cc}
		h_{\vex{k}} & \vec{\Xi}\cdot\vec{\mu}\\[1mm]
		\vec{\Xi}^{\dagger}\cdot\vec{\mu}	 & -\mu^yh_{-\vex{k}}^T\mu^y
	\end{array}\right],\\
	\vec{\Xi}=&\vec{\Delta}_{p,X}(-i\tau^y)\sigma^x+\vec{\Delta}_{p,Y}\tau^x(i\sigma^y)+\vec{\Delta}_{f}(-i\tau^y)+\vec{\Delta}_{\tilde{s}}(-i\tau^y)\sigma^z.
\end{align}

\section{Landau free energy}
We first assume that the order parameters are small. We use the identity $\ln\det A=\tr\ln A$. The free energy can be expressed by
\begin{align}
	F_{MFT}=&-T\tr\ln\left(\hat{G}^{-1}+\left[\begin{array}{cc}
		0& \vec{\Xi}\cdot\vec{\mu} \\[2mm]
		\vec{\Xi}^{\dagger}\cdot\vec{\mu} & 0
	\end{array}\right]\right)+\frac{A}{\tilde{U}}\left(|\vec{\Delta}_{\tilde{s}}|^2+|\vec{\Delta}_{p,X}|^2+|\vec{\Delta}_{p,Y}|^2+|\vec{\Delta}_{f}|^2\right)\\
	=&-T\left\{\tr\ln\hat{G}^{-1}-\tr\sum_{m=1}^{\infty}\frac{(-1)^m}{m}\left(\hat{G}\left[\begin{array}{cc}
		0& \vec{\Xi}\cdot\vec{\mu} \\[2mm]
		\vec{\Xi}^{\dagger}\cdot\vec{\mu} & 0
	\end{array}\right]\right)^m\right\}+\frac{A}{\tilde{U}}\left(|\vec{\Delta}_{\tilde{s}}|^2+|\vec{\Delta}_{p,X}|^2+|\vec{\Delta}_{p,Y}|^2+|\vec{\Delta}_{f}|^2\right)\\
	\label{Eq:F_MFT_expansion}\approx&\text{const}+T\tr\left[\hat{G}_+\left(\vec{\Xi}\cdot\vec{\mu}\right)\hat{G}_-\left(\vec{\Xi}^{\dagger}\cdot\vec{\mu}\right)\right]+\frac{A}{\tilde{U}}\left(|\vec{\Delta}_{\tilde{s}}|^2+|\vec{\Delta}_{p,X}|^2+|\vec{\Delta}_{p,Y}|^2+|\vec{\Delta}_{f}|^2\right)+O(|\Delta|^4)
\end{align}
where
\begin{align}
	\nonumber\hat{G}^{-1}(i\omega_n,\vex{k})=&\left[\begin{array}{cc}
		\hat{G}_+^{-1}(i\omega_n,\vex{k})	& 0\\
		0& 	\hat{G}_-^{-1}(i\omega_n,\vex{k})
	\end{array}\right]\\
	=&\left[\begin{array}{cc}
		(-i\omega_n-E_F)+v_F\left(\sigma^x\tau^zk_x+\sigma^yk_y\right)&  0\\[2mm]
		0 & (-i\omega_n+E_F) +v_F\left(\sigma^x\tau^zk_x-\sigma^yk_y\right)
	\end{array}\right],\\
	\label{Eq:G_+}	\hat{G}_+(i\omega_n,\vex{k})=&\frac{1}{(-i\omega_n-E_F)+v_F\left(\sigma^x\tau^zk_x+\sigma^yk_y\right)}=\frac{(i\omega_n+E_F)+v_F\left(\sigma^x\tau^zk_x+\sigma^yk_y\right)}{-(-i\omega_n-E_F)^2+v_F^2|\vex{k}|^2},\\
	\label{Eq:G_-}	\hat{G}_-(i\omega_n,\vex{k})=&\frac{1}{(-i\omega_n+E_F)+v_F\left(\sigma^x\tau^zk_x-\sigma^yk_y\right)}=\frac{(i\omega_n-E_F)+v_F\left(\sigma^x\tau^zk_x-\sigma^yk_y\right)}{-(-i\omega_n+E_F)^2+v_F^2|\vex{k}|^2}.
\end{align}
In Eq.~(\ref{Eq:F_MFT_expansion}), we assumed that the quartic order term in the order parameter is positive-definite. Thus, we can use the sign of the quadratic order terms to tell if a transition takes place (i.e., order parameters develop finite expectation values). This approach is consistent with the linearized gap equation approach which also treats the order parameters as perturbation.

Now, we evaluate the trace in the following.
\begin{align}
	&T\tr\left[\hat{G}_+\left(\vec{\Xi}\cdot\vec{\mu}\right)\hat{G}_-\left(\vec{\Xi}^{\dagger}\cdot\vec{\mu}\right)\right]
	=T\sum_{a=x,y,z}\sum_{\omega_n}\sum_{\vex{k}\in\frac{1}{2}\text{B.Z.}}\text{tr}\left[\begin{array}{cc}
		\hat{G}_+(i\omega_n,\vex{k})\Xi^a\hat{G}_-(i\omega_n,\vex{k})\left(\Xi^{\dagger}\right)^a
	\end{array}
	\right]\\
	=&T\sum_{\omega_n}\sum_{\vex{k}\in\frac{1}{2}\text{B.Z.}}\text{tr}\left[\begin{array}{cc}
		|\vec{\Delta}_{p,X}|^2\frac{(i\omega_n+E_F)+v_F\left(\sigma^x\tau^zk_x+\sigma^yk_y\right)}{-(-i\omega_n-E_F)^2+v_F^2|\vex{k}|^2}\left(-i\tau^y\sigma^x\right)\frac{(i\omega_n-E_F)+v_F\left(\sigma^x\tau^zk_x-\sigma^yk_y\right)}{-(-i\omega_n+E_F)^2+v_F^2|\vex{k}|^2}\left(i\tau^y\sigma^x\right)\\[2mm]
		+|\vec{\Delta}_{p,Y}|^2\frac{(i\omega_n+E_F)+v_F\left(\sigma^x\tau^zk_x+\sigma^yk_y\right)}{-(-i\omega_n-E_F)^2+v_F^2|\vex{k}|^2}\left(-i\tau^x\sigma^y\right)\frac{(i\omega_n-E_F)+v_F\left(\sigma^x\tau^zk_x-\sigma^yk_y\right)}{-(-i\omega_n+E_F)^2+v_F^2|\vex{k}|^2}\left(-i\tau^x\sigma^y\right)\\[2mm]
		+|\vec{\Delta}_{f}|^2\frac{(i\omega_n+E_F)+v_F\left(\sigma^x\tau^zk_x+\sigma^yk_y\right)}{-(-i\omega_n-E_F)^2+v_F^2|\vex{k}|^2}\left(-i\tau^y\right)\frac{(i\omega_n-E_F)+v_F\left(\sigma^x\tau^zk_x-\sigma^yk_y\right)}{-(-i\omega_n+E_F)^2+v_F^2|\vex{k}|^2}\left(i\tau^y\right)\\[2mm]
		+|\vec{\Delta}_{\tilde{s}}|^2\frac{(i\omega_n+E_F)+v_F\left(\sigma^x\tau^zk_x+\sigma^yk_y\right)}{-(-i\omega_n-E_F)^2+v_F^2|\vex{k}|^2}\left(-i\tau^y\sigma^z\right)\frac{(i\omega_n-E_F)+v_F\left(\sigma^x\tau^zk_x-\sigma^yk_y\right)}{-(-i\omega_n+E_F)^2+v_F^2|\vex{k}|^2}\left(i\tau^y\sigma^z\right)
	\end{array}
	\right]\\
	=&-8T\sum_{\omega_n}\sum_{\vex{k}\in\frac{1}{2}\text{B.Z.}}\left[\begin{array}{c}
		\frac{(\omega_n^2+E_F^2)\left(|\Delta_{p,X}|^2+|\Delta_{p,Y}|^2\right)}{\left[-(-i\omega_n+E_F)^2+v_F^2|\vex{k}|^2\right]\left[-(-i\omega_n-E_F)^2+v_F^2|\vex{k}|^2\right]}
		+\frac{(\omega_n^2+E_F^2+v_F^2|\vex{k}|^2)|\Delta_f|^2}{\left[-(-i\omega_n+E_F)^2+v_F^2|\vex{k}|^2\right]\left[-(-i\omega_n-E_F)^2+v_F^2|\vex{k}|^2\right]}\\[3mm]
		+\frac{(\omega_n^2+E_F^2-v_F^2|\vex{k}|^2)|\Delta_{\tilde{s}}|^2}{\left[-(-i\omega_n+E_F)^2+v_F^2|\vex{k}|^2\right]\left[-(-i\omega_n-E_F)^2+v_F^2|\vex{k}|^2\right]}
	\end{array}
	\right]\\
	=&-A\left[\xi(T)\left(|\Delta_{p,X}|^2+|\Delta_{p,Y}|^2\right)+\eta(T)|\Delta_f|^2+\zeta(T)|\Delta_{\tilde{s}}|^2\right],
\end{align}
where
\begin{align}
	\xi(T)=&8T\frac{1}{A}\sum_{\omega_n}\sum_{\vex{k}\in\frac{1}{2}\text{B.Z.}}\frac{\omega_n^2+E_F^2}{\left[-(-i\omega_n+E_F)^2+v_F^2|\vex{k}|^2\right]\left[-(-i\omega_n-E_F)^2+v_F^2|\vex{k}|^2\right]},\\
	\eta(T)=&8T\frac{1}{A}\sum_{\omega_n}\sum_{\vex{k}\in\frac{1}{2}\text{B.Z.}}\frac{\omega_n^2+E_F^2+v_F^2|\vex{k}|^2}{\left[-(-i\omega_n+E_F)^2+v_F^2|\vex{k}|^2\right]\left[-(-i\omega_n-E_F)^2+v_F^2|\vex{k}|^2\right]},\\
	\zeta(T)=&8T\frac{1}{A}\sum_{\omega_n}\sum_{\vex{k}\in\frac{1}{2}\text{B.Z.}}\frac{\omega_n^2+E_F^2-v_F^2|\vex{k}|^2}{\left[-(-i\omega_n+E_F)^2+v_F^2|\vex{k}|^2\right]\left[-(-i\omega_n-E_F)^2+v_F^2|\vex{k}|^2\right]}.
\end{align}

Clearly, $\eta(T)>\xi(T)>\zeta(T)$, implying that $f$-wave is dominating. To demonstrate this clearly, we first perform exact Matsubara summation and then evaluate the momentum space integral numerically. The integral expressions are as follows:
\begin{align}
	\label{Eq:eta_T}\eta(T)=&\frac{1}{2\pi}\int_0^{\Lambda} dk\,k \left[\frac{\tanh\left(\frac{\left|v_Fk-E_F\right|}{2T}\right)}{\left|v_Fk-E_F\right|}+\frac{\tanh\left(\frac{\left|v_Fk+E_F\right|}{2T}\right)}{\left|v_Fk+E_F\right|}\right],\\
	\label{Eq:xi_T}\xi(T)=&\frac{1}{2\pi}\int_0^{\Lambda} dk\, \left\{k \left[\frac{\tanh\left(\frac{\left|v_Fk-E_F\right|}{2T}\right)}{\left|v_Fk-E_F\right|}+\frac{\tanh\left(\frac{\left|v_Fk+E_F\right|}{2T}\right)}{\left|v_Fk+E_F\right|}\right]-\frac{v_F k^2}{2E_F} \left[\frac{\tanh\left(\frac{\left|v_Fk-E_F\right|}{2T}\right)}{\left|v_Fk-E_F\right|}-\frac{\tanh\left(\frac{\left|v_Fk+E_F\right|}{2T}\right)}{\left|v_Fk+E_F\right|}\right]\right\},\\
	\label{Eq:zeta_T}\zeta(T)=&\frac{1}{2\pi}\int_0^{\Lambda} dk\, \left\{k \left[\frac{\tanh\left(\frac{\left|v_Fk-E_F\right|}{2T}\right)}{\left|v_Fk-E_F\right|}+\frac{\tanh\left(\frac{\left|v_Fk+E_F\right|}{2T}\right)}{\left|v_Fk+E_F\right|}\right]-\frac{v_F k^2}{E_F} \left[\frac{\tanh\left(\frac{\left|v_Fk-E_F\right|}{2T}\right)}{\left|v_Fk-E_F\right|}-\frac{\tanh\left(\frac{\left|v_Fk+E_F\right|}{2T}\right)}{\left|v_Fk+E_F\right|}\right]\right\}.
\end{align}
where $\Lambda$ is the momentum cutoff.\\

With the above results and Eq.~(16) and Fig.~2 in the main text, we conclude that $f$-wave is the dominating instability. Technically, the effective BCS pairing interaction (after band projection) for the intra-sublattice pairings does not carry explicit angular dependent phase factor while the effective BCS pairing interaction for the inter-sublattice pairings contain angular dependent phase factor. As a result, the intra-sublattice pairing dominates over the inter-sublattice pairing. (Note that the sign change upon 180$^\circ$ rotation in the $f$-wave state is implemented by the valley Pauli matrix.)
In addition, the inter-sublattice pairing is typically energetically unfavorable in the presence of 
the asymmetry between A and B sites, which can be induced easily by either external potentials (e.g., a boron nitride substrate) or interlayer tunneling in the actual systems. These properties apply to graphene-based materials quite generally, and the realistic band structure is not needed for the understanding of the dominance of intra-sublattice pairing (i.e., $f$-wave).

\section{Free energy of f-wave superconductivity}

Now, we assume that $f$-wave is the dominating pairing channel and ignore the $s$-wave and $p$-wave pairings completely. The free energy is given by
\begin{align}
	F_{MFT}=-T\ln\det\left(-i\omega_n\hat{1}+\mathcal{H}_{\text{BdG},f}(\vex{k})\right)+\frac{A}{\tilde{U}}|\vec{\Delta}_f|^2,
\end{align}
where
\begin{align}
	\mathcal{H}_{\text{BdG},f}(\vex{k})=&\left[\begin{array}{cc}
		v_F\left(\sigma^x\tau^zk_x+\sigma^yk_y\right)-E_F& \vec{\Delta}_{f}\cdot\vec{\mu}(-i\tau^y) \\[2mm]
		\vec{\Delta}_{f}^*\cdot\vec{\mu}(i\tau^y) & -\mu^y\left(-v_F\sigma^x\tau^zk_x-v_F\sigma^yk_y-E_F\right)^T\mu^y
	\end{array}\right].
\end{align}

Now, we evaluate the determinant (but not on the $\omega_n$ and $\vex{k}$ spaces)
\begin{align}
	\nonumber&\det\left(-i\omega_n\hat{1}+\mathcal{H}_{\text{BdG},f}(\vex{k})\right)\\
	=&\det\left(\left[\begin{array}{cc}
		-i\omega_n+v_F\left(\sigma^x\tau^zk_x+\sigma^yk_y\right)-E_F& \vec{\Delta}_{f}\cdot\vec{\mu}(-i\tau^y) \\[2mm]
		\vec{\Delta}_{f}^*\cdot\vec{\mu}(i\tau^y) & -i\omega_n+v_F\left(\sigma^x\tau^zk_x-\sigma^yk_y\right)+E_F
	\end{array}\right]\right)\\
	\nonumber=&\det\left(-i\omega_n-E_F+v_F\left(\sigma^x\tau^zk_x+\sigma^yk_y\right)-\left(\vec{\Delta}_{f}\cdot\vec{\mu}\right)\left(\vec{\Delta}_{f}^*\cdot\vec{\mu}\right)\frac{1}{(-i\omega_n+E_F)+v_F\left(-\sigma^x\tau^zk_x-\sigma^yk_y\right)}\right)\\
	&\times\det\left((-i\omega_n+E_F)+v_F\left(\sigma^x\tau^zk_x-\sigma^yk_y\right)\right)\\
	\nonumber=&\det\left(\left[(-i\omega_n+E_F)+v_F\left(-\sigma^x\tau^zk_x-\sigma^yk_y\right)\right]\left[(-i\omega_n-E_F)+v_F\left(\sigma^xk_x+\sigma^y\tau^zk_y\right)\right]-\left(\vec{\Delta}_{f}\cdot\vec{\mu}\right)\left(\vec{\Delta}_{f}^*\cdot\vec{\mu}\right)\right)\\
	&\times\det\left((-i\omega_n+E_F)\hat{1}_8-v_F\left(\sigma^x\tau^zk_x-\sigma^yk_y\right)\right)/\det\left((-i\omega_n+E_F)+v_F\left(-\sigma^x\tau^zk_x-\sigma^yk_y\right)\right)\\
	=&\det\left(\left[-\omega_n^2-E_F^2-v_F^2|\vex{k}|^2-|\vec{\Delta}_f|^2\right]\hat{1}_8+2E_Fv_F\left(\sigma^x\tau^zk_x+\sigma^yk_y\right)-i\vec{\Delta}_f\times\vec{\Delta}_f^*\cdot\vec{\mu}
	\right)\\
	=&\left[\left(-\omega_n^2-v_F^2|\vex{k}|^2-E_F^2-|\vec{\Delta}_f|^2+\left|\vec{\Delta}_f^*\times\vec{\Delta}_f\right|\right)^2-4E_F^2v_F^2|\vex{k}^2|\right]^2\\
	&\times\left[\left(-\omega_n^2-v_F^2|\vex{k}|^2-E_F^2-|\vec{\Delta}_f|^2-\left|\vec{\Delta}_f^*\times\vec{\Delta}_f\right|\right)^2-4E_F^2v_F^2|\vex{k}^2|\right]^2,
\end{align}
where we have used the identity for block matrices
\begin{align}
	\det\left(\left[\begin{array}{cc}
		A & B\\
		C & D
	\end{array}\right]\right)=\det\left(A-BD^{-1}C\right)\det\left(D\right).
\end{align}

With the expression of $\det\left(-i\omega_n\hat{1}+\mathcal{H}_{\text{BdG},f}(\vex{k})\right)$, the free energy is expressed by
\begin{align}
	\label{Eq:F_MFT_fwave_omega_k}F_{MFT}=&-2T\sum_{\omega_n}\sum_{\vex{k}\in\frac{1}{2}\text{B.Z.}}\left\{\begin{array}{c}
		\ln\left[-\omega_n^2-\left(v_F|\vex{k}|-E_F\right)^2-\Delta_+^2\right]+\ln\left[-\omega_n^2-\left(v_F|\vex{k}|+E_F\right)^2-\Delta_+^2\right]\\[2mm]
		+\ln\left[-\omega_n^2-\left(v_F|\vex{k}|-E_F\right)^2-\Delta_-^2\right]+\ln\left[-\omega_n^2-\left(v_F|\vex{k}|+E_F\right)^2-\Delta_-^2\right]
	\end{array}\right\}+\frac{A}{\tilde{U}}|\vec{\Delta}_f|^2\\
	\label{Eq:F_MFT_fwave}=&-2TA\int\frac{d^2{\vex{k}}}{(2\pi)^2}\left\{\begin{array}{c}
		\ln\left[2\cosh\left(\frac{\sqrt{\left(v_F|\vex{k}|-E_F\right)^2+\Delta_+^2}}{2T}\right)\right]+\ln\left[2\cosh\left(\frac{\sqrt{\left(v_F|\vex{k}|+E_F\right)^2+\Delta_+^2}}{2T}\right)\right]\\[3mm]
		+\ln\left[2\cosh\left(\frac{\sqrt{\left(v_F|\vex{k}|-E_F\right)^2+\Delta_-^2}}{2T}\right)\right]+\ln\left[2\cosh\left(\frac{\sqrt{\left(v_F|\vex{k}|+E_F\right)^2+\Delta_-^2}}{2T}\right)\right]
	\end{array}
	\right\}+\frac{A}{\tilde{U}}|\vec{\Delta}_f|^2,
\end{align}
where $\Delta_{\pm}^2=|\vec{\Delta}_f|^2\pm\left|\vec{\Delta}_f^*\times\vec{\Delta}_f\right|$. To see the structure clearer, we derive the free energy density by expanding the order parameters in Eq.~(\ref{Eq:F_MFT_fwave}) in the following.
\begin{align}
	\mathcal{F}=&-2T\int\frac{d^2{\vex{k}}}{(2\pi)^2}\left\{\begin{array}{c}
		\ln\left[2\cosh\left(\frac{\sqrt{\left(v_F|\vex{k}|-E_F\right)^2+\Delta_+^2}}{2T}\right)\right]+\ln\left[2\cosh\left(\frac{\sqrt{\left(v_F|\vex{k}|+E_F\right)^2+\Delta_+^2}}{2T}\right)\right]\\[3mm]
		+\ln\left[2\cosh\left(\frac{\sqrt{\left(v_F|\vex{k}|-E_F\right)^2+\Delta_-^2}}{2T}\right)\right]+\ln\left[2\cosh\left(\frac{\sqrt{\left(v_F|\vex{k}|+E_F\right)^2+\Delta_-^2}}{2T}\right)\right]
	\end{array}
	\right\}+\frac{|\vec{\Delta}_f|^2}{\tilde{U}}\\
	\nonumber\approx&-2T\int\frac{d^2{\vex{k}}}{(2\pi)^2}\left\{\begin{array}{c}
		+\left[\frac{\tanh\left(\frac{\left|v_F|\vex{k}|+E_F\right|}{2T}\right)}{4T\left|v_F|\vex{k}|+E_F\right|}+\frac{\tanh\left(\frac{\left|v_F|\vex{k}|-E_F\right|}{2T}\right)}{4T\left|v_F|\vex{k}|-E_F\right|}\right]\left(\Delta_+^2+\Delta_-^2\right)\\[6mm]
		+\left[\frac{\text{sech}^2\left(\frac{\left|v_F|\vex{k}|+E_F\right|}{2 T}\right)}{16 T^2 \left|v_F|\vex{k}|+E_F\right|^2}-\frac{\tanh \left(\frac{\left|v_F|\vex{k}|+E_F\right|}{2 T}\right)}{8 T \left|v_F|\vex{k}|+E_F\right|^3}
		+\frac{\text{sech}^2\left(\frac{\left|v_F|\vex{k}|-E_F\right|}{2 T}\right)}{16 T^2 \left|v_F|\vex{k}|-E_F\right|^2}-\frac{\tanh \left(\frac{\left|v_F|\vex{k}|-E_F\right|}{2 T}\right)}{8 T \left|v_F|\vex{k}|-E_F\right|^3}\right]\frac{\Delta_+^4+\Delta_-^4}{2}
	\end{array}
	\right\}\\
	&+\frac{|\vec{\Delta}_f|^2}{\tilde{U}}+\text{const}\\
	=&\text{const}+\left[\frac{1}{\tilde{U}}-\eta(T)\right]|\vec{\Delta}_f|^2+a_4(T)\left[|\vec{\Delta}_f|^4+|\vec{\Delta}_f^*\times\vec{\Delta}_f|^2\right],
\end{align}
where
\begin{align}
	\label{Eq:eta_T_2}\eta(T)=&\int\frac{d^2{\vex{k}}}{(2\pi)^2}\left[\frac{\tanh\left(\frac{\left|v_F|\vex{k}|+E_F\right|}{2T}\right)}{\left|v_F|\vex{k}|+E_F\right|}+\frac{\tanh\left(\frac{\left|v_F|\vex{k}|-E_F\right|}{T}\right)}{\left|v_F|\vex{k}|-E_F\right|}\right]>0,\\
	a_4(T)=&-\int\frac{d^2{\vex{k}}}{(2\pi)^2}\left[\frac{\text{sech}^2\left(\frac{\left|v_F|\vex{k}|+E_F\right|}{2 T}\right)}{8 T \left|v_F|\vex{k}|+E_F\right|^2}-\frac{\tanh \left(\frac{\left|v_F|\vex{k}|+E_F\right|}{2 T}\right)}{4 \left|v_F|\vex{k}|+E_F\right|^3}
	+\frac{\text{sech}^2\left(\frac{\left|v_F|\vex{k}|-E_F\right|}{2 T}\right)}{8 T \left|v_F|\vex{k}|-E_F\right|^2}-\frac{\tanh \left(\frac{\left|v_F|\vex{k}|-E_F\right|}{2 T}\right)}{4 \left|v_F|\vex{k}|-E_F\right|^3}\right]>0.
\end{align}
The expression of $\eta(T)$ here is the same as the Eq.~(\ref{Eq:eta_T})
The results at quadratic order is consistent with our previous expansion with a Green function approach.\\

\section{$f$-wave superconductivity with a Zeeman splitting}

In the presence of a small in-plane magnetic field, we can incorporate the effect by adding a Zeeman splitting, which is model as $-B\mu^z$ without loss of generality. We evaluate the free energy by treating the order parameter perturbatively. The BdG Hamiltonian is given by
\begin{align}
	\mathcal{H}_{\text{BdG},f}(\vex{k})=&\left[\begin{array}{cc}
		v_F\left(\sigma^x\tau^zk_x+\sigma^yk_y\right)-B\mu^z-E_F& \vec{\Delta}_{f}\cdot\vec{\mu}(-i\tau^y) \\[2mm]
		\vec{\Delta}_{f}^*\cdot\vec{\mu}(i\tau^y) & -\mu^y\left(-v_F\sigma^x\tau^zk_x-v_F\sigma^yk_y-B\mu^z-E_F\right)^T\mu^y
	\end{array}\right].
\end{align}

\begin{align}
	F_{MFT}=&-T\tr\ln\left(\hat{G}_B^{-1}+\left[\begin{array}{cc}
		0& \vec{\Delta}_f\cdot\vec{\mu}(-i\tau^y) \\[2mm]
		\vec{\Delta}_f^*\cdot\vec{\mu}(i\tau^y) & 0
	\end{array}\right]\right)+\frac{A}{\tilde{U}}|\vec{\Delta}_f|^2\\
	\approx&\text{const}+T\tr\left[\hat{G}_{B+}\left(\vec{\Delta}_f\cdot\vec{\mu}\right)(-i\tau^y)\hat{G}_{B-}\left(\vec{\Delta}^{*}_f\cdot\vec{\mu}\right)(i\tau^y)\right]+\frac{A}{\tilde{U}}|\vec{\Delta}_f|^2+O(|\Delta|^4)
\end{align}
where
\begin{align}
	\nonumber\hat{G}_B^{-1}(i\omega_n,\vex{k})=&\left[\begin{array}{cc}
		\hat{G}_{B+}^{-1}(i\omega_n,\vex{k})	& 0\\
		0& 	\hat{G}_{B-}^{-1}(i\omega_n,\vex{k})
	\end{array}\right]\\
	\hat{G}_{h+}(i\omega_n,\vex{k})=&\frac{1}{(-i\omega_n-E_F)-B\mu^z+v_F\left(\sigma^x\tau^zk_x+\sigma^yk_y\right)}\approx \hat{G}_+(i\omega_n,\vex{k})+\hat{G}_+^2(i\omega_n,\vex{k})B\mu^z,\\
	\hat{G}_{h-}(i\omega_n,\vex{k})=&\frac{1}{(-i\omega_n+E_F)-B\mu^z-v_F\left(\sigma^x\tau^zk_x-\sigma^yk_y\right)}\approx \hat{G}_-(i\omega_n,\vex{k})+\hat{G}_-^2(i\omega_n,\vex{k})B\mu^z,
\end{align}
where $\hat{G}_{\pm}$ are given by Eqs.~(\ref{Eq:G_+}) and (\ref{Eq:G_-}). With the approximated Green functions, 
\begin{align}
	\nonumber&\tr\left[\hat{G}_{B+}\left(\vec{\Delta}_f\cdot\vec{\mu}\right)(-i\tau^y)\hat{G}_{B-}\left(\vec{\Delta}^{*}_f\cdot\vec{\mu}\right)(i\tau^y)\right]\\
	\nonumber\approx&\tr\left[\hat{G}_{+}\left(\vec{\Delta}_f\cdot\vec{\mu}\right)(-i\tau^y)\hat{G}_{-}\left(\vec{\Delta}^{*}_f\cdot\vec{\mu}\right)(i\tau^y)\right]\\
	\nonumber&+\tr\left[\hat{G}_{+}\left(\vec{\Delta}_f\cdot\vec{\mu}\right)(-i\tau^y)\hat{G}_{-}^2B\mu^z\left(\vec{\Delta}^{*}_f\cdot\vec{\mu}\right)(i\tau^y)\right]+\tr\left[\hat{G}_{+}^2B\mu^z\left(\vec{\Delta}_f\cdot\vec{\mu}\right)(-i\tau^y)\hat{G}_{-}\left(\vec{\Delta}^{*}_f\cdot\vec{\mu}\right)(i\tau^y)\right]\\
	&+\tr\left[\hat{G}_{+}^2B\mu^z\left(\vec{\Delta}_f\cdot\vec{\mu}\right)(-i\tau^y)\hat{G}_{-}^2B\mu^z\left(\vec{\Delta}^{*}_f\cdot\vec{\mu}\right)(i\tau^y)\right]+O(B^3).
\end{align}

The linear in $B$ term is given by
\begin{align}
	\nonumber&T\tr\left[\hat{G}_{+}\left(\vec{\Delta}_f\cdot\vec{\mu}\right)(-i\tau^y)\hat{G}_{-}^2B\mu^z\left(\vec{\Delta}^{*}_f\cdot\vec{\mu}\right)(i\tau^y)\right]
	+\tr\left[\hat{G}_{+}^2B\mu^z\left(\vec{\Delta}_f\cdot\vec{\mu}\right)(-i\tau^y)\hat{G}_{-}\left(\vec{\Delta}^{*}_f\cdot\vec{\mu}\right)(i\tau^y)\right]\\
	\nonumber=&BT\tr\left[\left\{\frac{(i\omega_n+E_F)+v_F\left(\sigma^x\tau^zk_x+\sigma^yk_y\right)}{-(-i\omega_n-E_F)^2+v_F^2|\vex{k}|^2}\right\}\left\{\frac{(i\omega_n-E_F)+v_F\left(-\sigma^x\tau^zk_x-\sigma^yk_y\right)}{-(-i\omega_n+E_F)^2+v_F^2|\vex{k}|^2}\right\}^2\mu^z\left(\vec{\Delta}^{*}_f\cdot\vec{\mu}\right)\left(\vec{\Delta}_f\cdot\vec{\mu}\right)\right]\\
	&+BT\tr\left[\left\{\frac{(i\omega_n+E_F)+v_F\left(\sigma^x\tau^zk_x+\sigma^yk_y\right)}{-(-i\omega_n-E_F)^2+v_F^2|\vex{k}|^2}\right\}^2\left\{\frac{(i\omega_n-E_F)+v_F\left(-\sigma^x\tau^zk_x-\sigma^yk_y\right)}{-(-i\omega_n+E_F)^2+v_F^2|\vex{k}|^2}\right\}\mu^z\left(\vec{\Delta}_f\cdot\vec{\mu}\right)\left(\vec{\Delta}^{*}_f\cdot\vec{\mu}\right)\right]\\
	=&-iBT\left(\vec{\Delta}_f\times\vec{\Delta}_f^*\right)\cdot\hat{z}\tr\left[\begin{array}{c}
		\left\{\frac{(i\omega_n+E_F)+v_F\left(\sigma^x\tau^zk_x+\sigma^yk_y\right)}{-(-i\omega_n-E_F)^2+v_F^2|\vex{k}|^2}\right\}\left\{\frac{(i\omega_n-E_F)+v_F\left(-\sigma^x\tau^zk_x-\sigma^yk_y\right)}{-(-i\omega_n+E_F)^2+v_F^2|\vex{k}|^2}\right\}^2\\[2mm]
		-\left\{\frac{(i\omega_n+E_F)+v_F\left(\sigma^x\tau^zk_x+\sigma^yk_y\right)}{-(-i\omega_n-E_F)^2+v_F^2|\vex{k}|^2}\right\}^2\left\{\frac{(i\omega_n-E_F)+v_F\left(-\sigma^x\tau^zk_x-\sigma^yk_y\right)}{-(-i\omega_n+E_F)^2+v_F^2|\vex{k}|^2}\right\}
	\end{array}
	\right]\\
	\equiv&-ia_1(T)BA\left(\vec{\Delta}_f\times\vec{\Delta}_f^*\right)\cdot\hat{z}.
\end{align}
The sign of $a_1(T)$ is hard to see without evaluating the integrals. However, we known that $a_1(T)>0$ based on physical arguments.\\

The quadratic order in $B$ is given by
\begin{align}
	\nonumber&T\tr\left[\hat{G}_{+}^2B\mu^z\left(\vec{\Delta}_f\cdot\vec{\mu}\right)(-i\tau^y)\hat{G}_{-}^2B\mu^z\left(\vec{\Delta}^{*}_f\cdot\vec{\mu}\right)(i\tau^y)\right]\\
	=&B^2T\tr\left[\left\{\frac{(i\omega_n+E_F)+v_F\left(\sigma^x\tau^zk_x+\sigma^yk_y\right)}{-(-i\omega_n-E_F)^2+v_F^2|\vex{k}|^2}\right\}^2\left\{\frac{(i\omega_n-E_F)+v_F\left(-\sigma^x\tau^zk_x-\sigma^yk_y\right)}{-(-i\omega_n+E_F)^2+v_F^2|\vex{k}|^2}\right\}^2\mu^z\left(\vec{\Delta}_f\cdot\vec{\mu}\right)\mu^z\left(\vec{\Delta}^{*}_f\cdot\vec{\mu}\right)\right]\\
	=&B^2T\left(-|\Delta_f^x|^2-|\Delta_f^y|^2+|\Delta_f^z|^2\right)\tr\left[\left\{\frac{(i\omega_n+E_F)+v_F\left(\sigma^x\tau^zk_x+\sigma^yk_y\right)}{-(-i\omega_n-E_F)^2+v_F^2|\vex{k}|^2}\right\}^2\left\{\frac{(i\omega_n-E_F)+v_F\left(-\sigma^x\tau^zk_x-\sigma^yk_y\right)}{-(-i\omega_n+E_F)^2+v_F^2|\vex{k}|^2}\right\}^2\right]\\
	\equiv& \delta\eta(T) B^2A\left(-|\Delta_f^x|^2-|\Delta_f^y|^2+|\Delta_f^z|^2\right),
\end{align}
where $\delta \eta(T)>0$.

With both the $a_1(T)$ and $\delta\eta(T)$ terms, the free energy density upto quadratic order can be expressed by
\begin{align} \mathcal{F}=&\text{const}+\left[\frac{1}{\tilde{U}}-\eta(T)-B^2\delta\eta(T)\right]\left(|\Delta_f^x|^2+|\Delta_f^y|^2\right)+\left[\frac{1}{\tilde{U}}-\eta(T)+B^2\delta\eta(T)\right]|\Delta_f^z|^2-ia_1(T)B\left(\vec{\Delta}_f\times\vec{\Delta}_f^*\right)\cdot\hat{z}.
\end{align}
In the main text, we have defined $\eta_x(T)$, $\eta_y(T)$, and $\eta_z(T)$. In the presence of a Zeeman splitting, $\eta_x(T)=\eta_y(T)=\eta(T)+B^2\delta\eta(T)$
and $\eta_z(T)=\eta(T)-B^2\delta\eta(T)$. Thus, we conclude that $\Delta_f^x$ and $\Delta_f^y$ are favored based on the form of the free energy.

\section{Experimental setup for $f$-wave spin-triplet pairing}

\begin{figure}[h!]
	\includegraphics[width=0.5\textwidth]{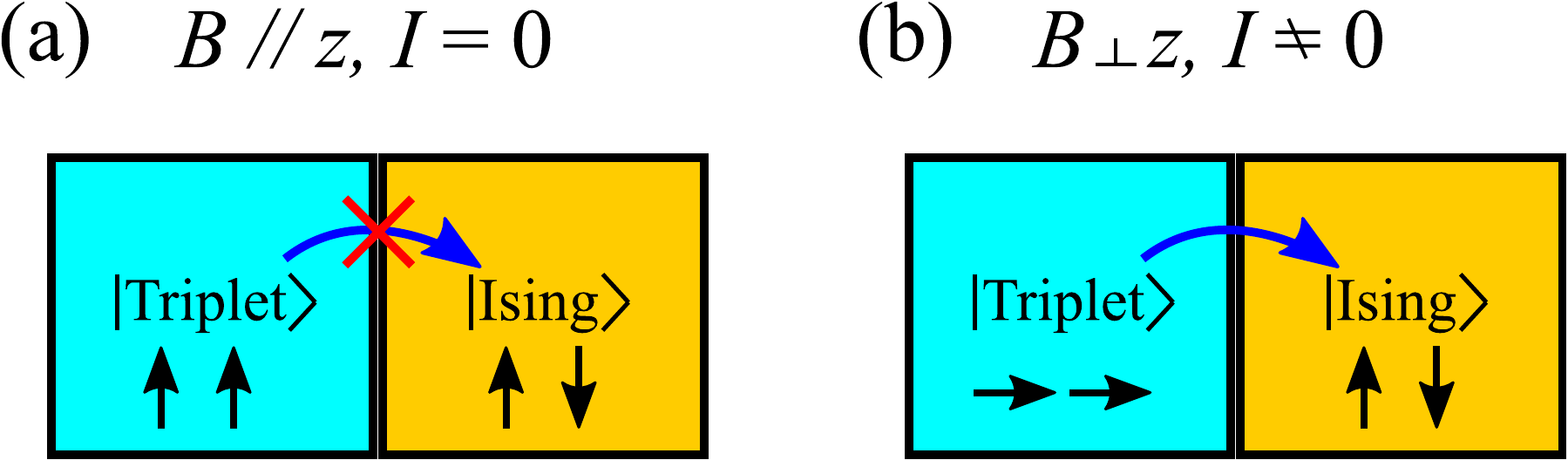}
	\caption{Josephson junction made of a spin-triplet superconductor and an Ising superconductor. The Josephson tunneling can be controlled by the mall external magnetic field which polarize the triplet Cooper pairs. (a) There is no Josephson tunneling when the applied field is out-of-plane. (b) When the applied field is in-plane, a finite Josephson current can occur. 
	}
	\label{Fig:Josephson}
\end{figure}

In this section, we propose a Josephson junction setup that can identify the $f$-wave spin-triplet superconductivity. There are two main ingredients: (a) Identification of spin-triplet and (b) identification of $f$-wave. The ideas are summarized in Fig.~\ref{Fig:Josephson}.

\subsection{Identification of spin-triplet pairing}

Since spin-singlet and spin-triplet are mutually orthogonal, we cannot use the conventional BCS type of superconductor to make the Josephson junctions. One possibility is to find a superconducting material with spin-orbit coupling which hybridize the singlet and triplet pairing.
To this end, the NbSe$_2$ is an established example of the ``Ising'' superconductor \cite{Xi2016ising,Kim2017strong}, whose spin wavefunction is given by
\begin{align}
	|\text{Ising}\rangle=&|\uparrow\rangle_{\text{K}}|\downarrow\rangle_{\text{K}'}=\frac{1}{2}\left[\left(|\uparrow\rangle_{\text{K}}|\downarrow\rangle_{\text{K}'}+|\downarrow\rangle_{\text{K}}|\uparrow\rangle_{\text{K}'}\right)+\left(|\uparrow\rangle_{\text{K}}|\downarrow\rangle_{\text{K}'}-|\downarrow\rangle_{\text{K}}|\uparrow\rangle_{\text{K}'}\right)\right]\\
	\label{Ising_spin}=&\frac{1}{\sqrt{2}}\left(|\text{triplet},S^z=0\rangle+|\text{singlet}\rangle\right).
\end{align}
Thus, the Ising Cooper pair can be seen as a mixture of spin-singlet and spin triplet Cooper pairs.

Our predicted $f$-wave spin-triplet superconductor is partially polarized in the presence of the in-plane magnetic field. Without loss of generality, we assume the magnetic field is along $x$ direction. The spin wavefunction of the dominating Cooper pair is given by
\begin{align}
	|\Phi_f\rangle=&|\rightarrow\rangle_{\text{K}}|\rightarrow\rangle_{\text{K}'}=\frac{1}{2}\left(|\uparrow\rangle_{\text{K}}+|\downarrow\rangle_{\text{K}}\right)
	\left(|\uparrow\rangle_{\text{K}'}+|\downarrow\rangle_{\text{K}'}\right)=\frac{1}{2}\left[|\uparrow\rangle_{\text{K}}|\uparrow\rangle_{\text{K}'}+|\downarrow\rangle_{\text{K}}|\downarrow\rangle_{\text{K}'}+|\uparrow\rangle_{\text{K}}|\downarrow\rangle_{\text{K}'}+|\downarrow\rangle_{\text{K}}|\uparrow\rangle_{\text{K}'}\right]\\
	\label{f_spin}=&\frac{1}{2}|\text{triplet},S^z=1\rangle+\frac{1}{2}|\text{triplet},S^z=-1\rangle+\frac{1}{\sqrt{2}}|\text{triplet},S^z=0\rangle.
\end{align}
With Eqs.~(\ref{Ising_spin}) and (\ref{f_spin}), the overlap of the wavefunctions $\langle \text{Ising}|\Phi_f \rangle=1/2$, suggesting that a finite Josephson tunneling between a spin-triplet superconductor and an Ising superconductor. By contrast, we also consider the small out-of-plane magnetic field (i.e., the orbital effect is ignored). In the presence of a magnetic field along $z$ direction, the spin wavefunction of the dominating Cooper pair is given by $|\Phi_f\rangle=|\uparrow\rangle_{\text{K}}|\uparrow\rangle_{\text{K}'}$, which does not overlap with the Ising pairing wavefunction.\\ 

\begin{figure}[h]
	\includegraphics[width=0.3\textwidth]{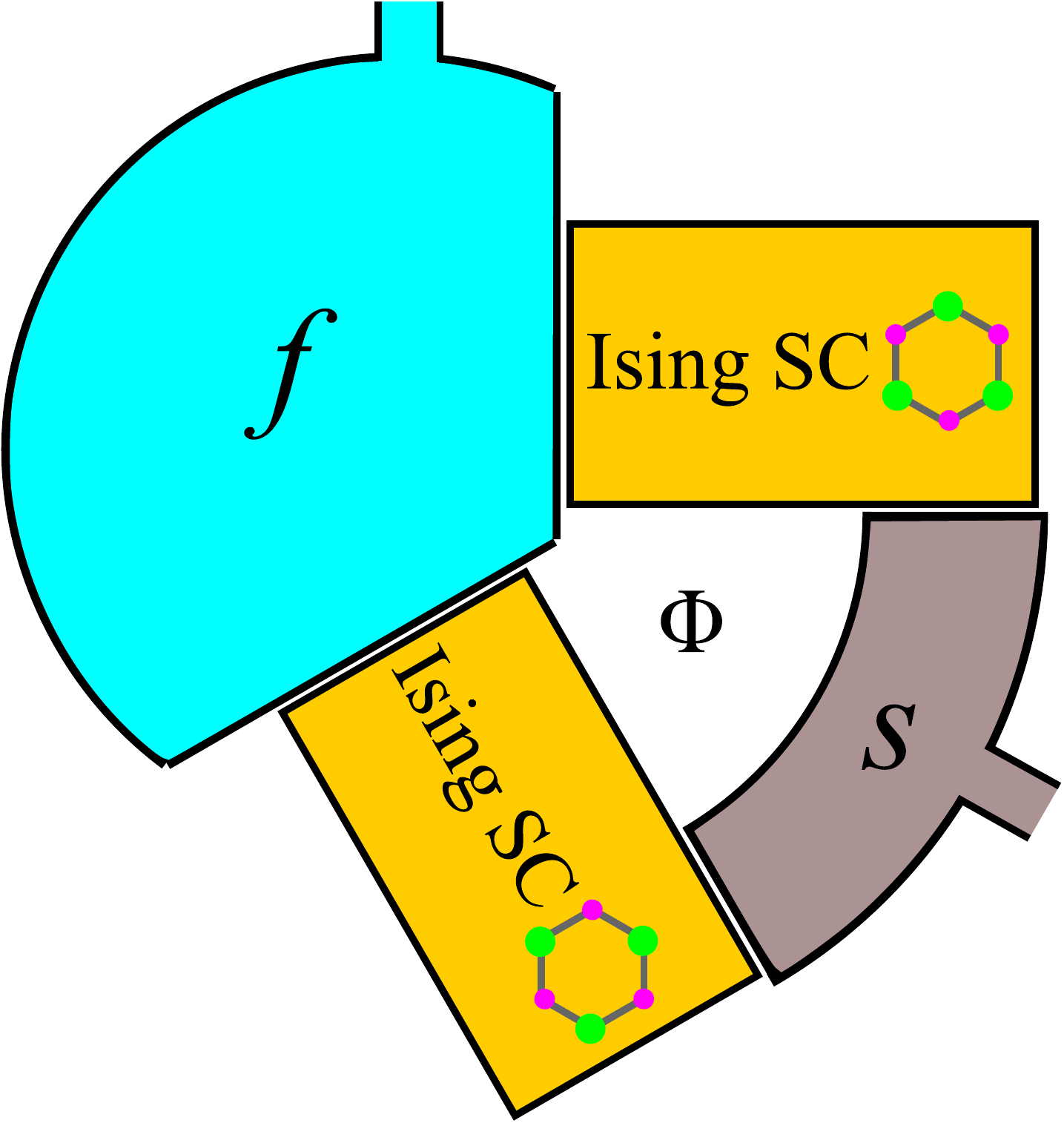}
	\caption{Hybrid corner junction setup. We consider two Ising superconductors (yellow region) connected to the $f$-wave superconductor (light blue region). The lower Ising superconductor is physically rotated by $\pi/3$ relative to the upper Ising superconductor. We also use an $s$-wave superconductor (pale brown) to connect the two Ising superconductors so that a close loop is formed. Since the $f$-wave pairing symmetry contributes to a relative $-1$ under a $\pi/3$ rotation, the flux must be shifted by half of the flux quantum. The $\pi$ phase shift can be detected by the current-voltage relation experimentally.
	}
	\label{Fig:Corner}
\end{figure}

The spin-triplet pairing can be confirmed by Josephson junction with an Ising superconductor and a magnetic field. The ideas are summarized in Fig.~\ref{Fig:Josephson}(a) and (b).
With a small in-plane magnetic field, the Josephson tunneling is finite and the overlap $\langle \text{Ising}|\Phi_f \rangle$ is $1/2$. With a small out-of-plane magnetic field, the Josephson current is suppressed to zero. These should be tested by the existing experimental techniques.
It is worth mentioning that the superconductivity in NbSe$_2$ can also persist in the presence of a large in-plane magnetic field \cite{Xi2016ising}. Thus, the effect of the \textit{small} in-plane magnetic field can be ignored in NbSe$_2$.

\subsection{$f$-wave symmetry}

We can use a hybrid corner junction setup (Fig.~\ref{Fig:Corner}) to confirm the $f$-wave symmetry in MATTG superconductivity. This setup requires both an Ising superconductors (such as NbSe$_2$) and an $s$-wave superconductor (such as Pb).
One subtle issue is that the Ising Cooper pair is a linear combination of singlet and triplet states, so there can be a nontrivial phase factor upon rotation. To avoid the nontrivial phase contributions from the Ising superconductor, the two NbSe$_2$ devices are rotated relatively by $\pi/3$ as illustrated in the Fig.~\ref{Fig:Corner}. Therefore, the phase accumulation in the two MATTG-NbSe$_2$ junctions is purely from the pairing symmetry in the superconducting state in MATTG.
Since the $f$-wave order contributes to a minus sign upon the $\pi/3$ rotation, the two MATTG-NbSe$_2$ junctions (with spin-triplet Cooper pair tunneling) pick up a $\pi$ phase shift in total. On the other hand, there is no nontrivial phase contribution in the NbSe$_2$-Pb junctions (with spin-singlet Cooper pair tunneling) as the singlet sector is invariant under a $\pi$ rotation.
Therefore, the $f$-wave pairing symmetry corresponds to a $\pi$ phase shift in corner junction setup, which can be detected through the current-voltage relation.
The same setup can be used to detect other pairing symmetry such as $s$-, $p$-, and $d$-wave pairing, corresponding to $0$, $\pm 2\pi/3$, and $\pm 4\pi/3$ phase shifts respectively.


\begin{thebibliography}{59}
\bibitem[{Sta()}]{Staggered}
	\bibinfo{note}{This pairing can be derived from the $f$-wave state as the
		difference is a $\sigma^z$ in the expression. The $\mathcal{C}_{2z}$ and
		$\mathcal{C}_{3z}$ operations are consistent with $s$-wave, but this is not
		the conventional $s$-wave state due to the staggered sublattice structure.}
\bibitem[{TR()}]{TR}
	\bibinfo{note}{Formally, $\hat{\Pi}=\tau^x\mu^y$, corresponding to
		$\mathcal{T}^2=-1$, is the physical time-reversal symmetry. The existence of
		$\mathcal{T}^2=1$ time-reversal operation is resulted from the
		spin-conservation. Thus, the irreducible diagonal block Hamiltonian should be
		viewed as a system with effectively $\mathcal{T}^2=1$.}
\bibitem[{DP()}]{DP}
	\bibinfo{note}{We note that Eqs.~(\ref{Eq:Delta_0}) and (\ref{Eq:T_c}) fail
		when $E_F=0$ (Dirac point), which is in fact an unstable non-Fermi-liquid
		fixed point, indicating the breakdown of BCS theory without a finite Fermi
		surface.}
\end{thebibliography}
\end{document}